\documentclass[twocolumn,showpacs,preprintnumbers,prl,amsmath,amssymb,floatfix]{revtex4}
\usepackage{epsfig}
\begin{document}
%
%

\title{Ferromagnetic Luttinger Liquids}
%
\author{Lorenz Bartosch, Marcus Kollar, and Peter Kopietz}
\affiliation{Institut f\"{u}r Theoretische Physik, Universit\"{a}t Frankfurt,
Robert-Mayer-Strasse 8, 60054 Frankfurt, Germany}
\date{July 12, 2002}
\begin{abstract}
We study weak  itinerant ferromagnetism  in one-dimensional 
Fermi systems using perturbation theory and bosonization.
We find that longitudinal spin fluctuations propagate
ballistically with  velocity  $v_m \ll v_F$, where $v_F$ is the Fermi velocity.
This leads to a large  anomalous dimension in the spin-channel
and strong algebraic singularities
in the single-particle spectral function and in the
transverse structure factor for momentum transfers
$q \approx 2 \Delta / v_F$,   where $2 \Delta$ is the exchange splitting.

\end{abstract}
\maketitle

Recently several authors presented conductance
measurements in ultra low-disorder semiconductor quantum wires and
suggested that an unusual feature in the range $0.5 - 0.7 \times 2 e^2
/h$ of conductance can be explained in terms of spontaneous
ferromagnetism \cite{Thomas96,Reilly02,PhysicsToday02}.  At first sight this
interpretation seems to contradict the Lieb-Mattis theorem
\cite{Lieb62}, which rules out magnetized ground states for electrons
moving on a line, as well as for one-band lattice models in one dimension
($1d$) with nearest-neighbor hopping and interactions involving densities.
However, there is no fundamental principle that forbids ferromagnetic
ground states in quasi $1d$ systems with finite width or 
one-band lattice models in $1d$ with
more general hoppings.  Indeed, numerical studies \cite{Daul98} show that the
ground state of the $1d$ Hubbard model with hopping between nearest
and next-nearest neighbors can be ferromagnetic in a substantial range of
densities and on-site interactions $U$.
Clearly,  the precise form of the energy dispersion $\epsilon_k$ plays an important
role in stabilizing ferromagnetism \cite{Moriya85,Mielke93,Vollhardt01}.
In principle, it should therefore be possible
to design metallic systems with ferromagnetic ground states
by properly adjusting the hopping integrals between the relevant orbitals. 
A promising class of $1d$ materials where this might be achieved
are certain types of organic polymers \cite{Arita02}, whose
molecular structure
can be designed in a controlled manner in the laboratory.
Motivated by these new developments, in this work we shall use a combination
of perturbation theory and bosonization to derive 
some  physical properties of itinerant ferromagnets in $1d$.

Let us briefly consider this problem from a
renormalization group (RG) point of view.
The usual RG approach to $1d$ metals is based
on the assumption that their long-wavelength and low-energy properties 
are determined by wavevectors $k$ in the vicinity of the 
Fermi wavevectors  $\pm k_F$. Given a general energy dispersion $\epsilon_k$, it therefore seems
reasonable to expand   for $k $ close to $ k_F$
\begin{equation}
 \epsilon_k =  \epsilon_{ k_F } +   v_F ( k -  k_F ) +
 \frac{  ( k -  k_F )^2  }{ 2 m^{\ast} } +
  \frac{ \lambda }{6}   ( k -  k_F )^3  + \ldots
 \; ,
 \label{eq:energydispersion}
 \end{equation}
and similarly for $ k \approx - k_F$.
By power counting, the Fermi velocity $v_F$ is a marginal coupling,
while the inverse effective mass $1 / m^{\ast}$ and the cubic parameter $\lambda$ are 
irrelevant in the RG sense. 
In the field-theoretical formulation of the RG \cite{Solyom79}, these irrelevant couplings are
simply ignored.
However, as shown below,  the cubic term in Eq.\ (\ref{eq:energydispersion})
is crucial to stabilize a ferromagnetic ground state in $1d$, so that
a proper RG treatment of itinerant ferromagnetism should include also 
the irrelevant couplings associated with band
curvature effects.
Therefore methods which cannot properly handle
these couplings,  such as the field-theoretical RG  \cite{Solyom79} or  bosonization,  
lose much of their power.
Nevertheless, as shown below,  in certain regimes bosonization is still useful to 
obtain nonperturbative results for correlation functions.

We consider the following  Hamiltonian describing interacting electrons 
on a $1d$ lattice with length $L$,
 \begin{equation}
 \hat{H}   =  
 \sum_{k  \sigma}  \epsilon_k 
 \hat{c}^{\dagger}_{ k \sigma} \hat{c}^{\phantom{\dagger}}_{ k \sigma } 
  + 
 \frac{1}{2 L} \sum_{q, ij}   
 f_{ij}\hat{\rho}_i (-q ) \hat{\rho}_{j} ( q )
 \; ,
 \label{eq:Hamiltonian}
 \end{equation}
where
$\hat{c}^{\dagger}_{ k \sigma}$ and $\hat{c}^{\phantom{\dagger}}_{k \sigma}$ are creation and annihilation operators for electrons with momentum $k$ and
spin $\sigma$. The labels $i$ and $j$ assume  values in 
$\{ n,m \}$, where $n$ corresponds to  the charge
density $\hat{\rho}_n ( q ) = \sum_{ k \sigma}  \hat{c}^{\dagger}_{k \sigma} 
\hat{c}^{\phantom{\dagger}}_{k+q \sigma}$,  and $m$  denotes the spin density 
$  \hat{\rho}_m ( q ) = \sum_{ k \sigma} \sigma \hat{c}^{\dagger}_{k \sigma} 
\hat{c}^{\phantom{\dagger}}_{k+q \sigma}$.
To discuss {\it{spontaneous}} symmetry breaking we should start from a
spin-rotationally invariant $\hat{H}$, which constrains the bare
$f_{ij}$ to satisfy $f_{nm} = f_{mn} =0$ and precludes any momentum-dependence of 
$f_m \equiv f_{mm}$. We also take $f_n \equiv f_{nn}$ to be momentum-independent \cite{footnoteHubbard}. 

As a first step, we study the ferromagnetic instability  within Hartree-Fock theory.
Adding and subtracting the counterterm  $ {\Delta}_{\sigma}   ( m ) = f_n n + \sigma f_m m $,
where $ n = \langle \hat{\rho}_n ( 0) \rangle/L$ is the density and
$m = \langle \hat{\rho}_m ( 0) \rangle/L$ is the spin density,
we may write
$ \hat{H}  = \hat{H}_0 + \hat{H}_{1}$, with 
 \begin{equation}
 \hat{H}_0 - \mu \hat{N} = \sum_{k  \sigma} 
 \xi_{ k \sigma}
 \hat{c}^{\dagger}_{ k \sigma} \hat{c}^{\phantom{\dagger}}_{ k \sigma }  - 
\frac{L}{2} [ f_n n^2 + f_m m^2 ] 
 \; ,
 \end{equation}
and
$ \hat{H}_{1} =
 (2 L)^{-1} \sum_{q,i} f_{i}  
  \delta \hat{\rho}_i ( -q )  
  \delta \hat{\rho}_i ( q ).$ 
Here 
 $\xi_{ k \sigma} =
 \epsilon_k - \mu + {\Delta}_{\sigma} ( m )$ is the Hartree-Fock  energy, 
$\delta \hat{\rho}_i ( q ) = \hat{\rho}_i ( q ) - \delta_{q,0}  \langle \hat{\rho}_i (0) \rangle$, and
$\hat{N} = \hat{\rho}_n ( 0 )$.
In the ferromagnetic state the Fermi wavevectors
$k_{\sigma}$ and velocities $v_{\sigma}$ are  defined  by
 $\epsilon_{ k_{ \sigma}} - \mu +  {\Delta}_{\sigma} ( m ) = 0 $ and 
$  v_{ \sigma} = 
 \left.  \partial \epsilon_k / \partial k \right|_{  k_{ \sigma}}$, while
in the normal state
 $ \epsilon_{k_F} - \mu + {\Delta}_{\sigma} ( 0 ) =0$ and
$v_F =   \partial \epsilon_k  / \partial k |_{  k_{F}}$.
Hence 
 $\epsilon_{ k_{\sigma} } - \epsilon_{ k_F } + f_n \delta n =  \sigma \Delta$
where $ \Delta = - f_m m$ and $\delta n = n (m ) - n ( 0 )$.
For convenience we keep the chemical potential $\mu$ constant, so that
the density $n$ is a function of  $m$. 
The two equations
$\epsilon_{ k_{\sigma} } - \epsilon_{ k_F } + f_n \delta n =  \sigma \Delta$,
$\sigma = \pm 1$,
together with the self-consistency conditions
$m = \pi^{-1} ( k_{\uparrow} - k_{\downarrow} )$ and
$\delta n = \pi^{-1} ( k_{\uparrow} + k_{\downarrow} - 2 k_F)$
fix the four quantities
$ k_{ \uparrow}$, $k_{  \downarrow}$, $\delta n$, and  $m$.

Throughout this work we shall assume  $m \ll n$ (weak ferromagnetism).
The low-energy properties are then
determined by wavevectors  in the vicinity of the Fermi surface,
as discussed in the classic work by Dzyaloshinski\u{i} and Kondratenko \cite{Dzyaloshinskii76}.
Hence we may expand $\epsilon_k$ around
$\pm k_F$. To leading order, it is sufficient to truncate the expansion at the
third order, see Eq.\ (\ref{eq:energydispersion}).
Keeping in mind that $\pi m \ll k_F$ and defining
$q_m = \Delta / v_F$ we obtain
\begin{eqnarray}
 k_{  \sigma} - k_F  & = &  \sigma q_m - \frac{ \lambda_1  q_m^2}{2 ( 1 + F_0)}
 - 2 \sigma A q_m^3
+ \ldots
 \label{eq:fksigmasol}
 \; ,
 \end{eqnarray}
where $A = \frac{1}{12} (   \lambda_2 - \frac{3 \lambda_1^2}{ 1 + F_0 } )$, with
$ \lambda_1 = 1 / ( m^{\ast} v_F )$,  $ \lambda_2 = \lambda / v_F$, and
$F_0 = 2 f_n /  \pi v_F $.
The Fermi velocities are
 \begin{equation}
 v_{  \sigma}   / v_F 
  = 1 +  
 \sigma  \lambda_1 q_m + \frac{1}{2} \left(  \lambda_2 - \frac{\lambda_1^2}{1 + F_0}  
 \right)  q_m^2
 + \ldots  
\label{eq:vFsigmasol}
 \; .
 \end{equation}
Substituting Eq.\ (\ref{eq:fksigmasol}) into $ m = \pi^{-1} ( k_{\uparrow} -
k_{\downarrow} )$, it is easy to see that,
besides the solution $ m=0$, there is a nontrivial solution 
$ \pi m_0 = [ 2 (  I_0 -1 ) /   ( I_0^3 A)  ]^{1/2}$, provided the radicand is positive.
Here
$ I_0 = - 2 f_m / \pi v_F $ is the dimensionless Stoner parameter \cite{Moriya85}.
To see whether the  solution $m_0$ is stable,  
we consider 
the energy change $\delta \Omega_0 ( m )
= \Omega_0 ( m ) - \Omega_0 ( 0)$
due to a finite value of $m$, where
$ \Omega_0 ( m ) =  \langle \hat{H}_0 - \mu \hat{N} \rangle$.
We obtain 
 \begin{equation}
 \delta \Omega_0 ( m )   
  =   \frac{L v_F}{ 4 \pi }
 \biggl[ -   I_0 ( I_0 -1 )  ( \pi m )^2
  +   \frac{A}{4} I_0^4  ( \pi m )^4
 + \ldots
 \biggr]
 \; .
 \label{eq:hartreegain}
 \end{equation}
Obviously, 
a necessary condition for $m_0$ to represent a minimum of $\Omega_0 (m)$ is
$ A \geq 0$.
In addition, the square root   $[ 2 (  I_0 -1 ) /   ( I_0^3 A)  ]^{1/2}$ 
is only real if
either $I_0 < 0$ or
$ I_0 > 1$. For consistency, we should also require that 
$ \pi m_0 \ll k_F$ and that
the band-structure is such that
the higher order corrections in Eq.\ (\ref{eq:hartreegain}) are small.
For some special  form of $\epsilon_k$ it should be possible
to  satisfy these conditions 
even for small negative $I_0$
provided $k_F^2 A \gg \pi^2 | I_0 |^{-3}$.
Here we shall not further consider this case, but 
focus instead on the  regime close to the Stoner threshold, where
 $ I_0 $ is slightly larger than unity. The distance
from the critical point is then measured by the small parameter
$ \delta_0  \equiv   2 (I_0 -1)/I_0  $. 
Interestingly, the numerical results of Ref.~\onlinecite{Daul98} indeed
show a critical $I_0$ of order unity for not too large
densities, which suggests that even in $1d$ the Stoner criterion can be a
reasonable estimate for the ferromagnetic instability.

%

For simplicity we now set
$ f_n = - f_m  = f_0 > 0$, corresponding to  
a repulsive Hubbard on-site interaction \cite{footnoteHubbard}.
Note that close to the phase transition
$I_0   = F_0 = 1 + {\mathcal{O}} ( \delta_0 )$.
Let us first consider the density-density ($\chi_{nn}$) and the
longitudinal spin-spin ($\chi_{mm}$) correlation functions.
Within the 
random-phase approximation  (RPA) we obtain
 \begin{subequations}
 \begin{eqnarray}
 \chi_{nn}^{\rm RPA} ( q , i \omega )  & = & [ \chi_{\uparrow \uparrow}^0 +
 \chi_{\downarrow \downarrow}^0  - 4 f_0  \chi_{\uparrow \uparrow}^{0}
   \chi_{\downarrow \downarrow}^{0} ] / { D }  
 \; ,
 \label{eq:chinn}
 \\
 \chi_{mm}^{\rm RPA}  ( q , i \omega ) & = & [ \chi_{\uparrow \uparrow}^0 +
 \chi_{\downarrow \downarrow}^0  + 4 f_0  \chi_{\uparrow \uparrow}^{0}
   \chi_{\downarrow \downarrow}^{0} ] / { D }  
 \; ,
 \label{eq:chimm}
 \end{eqnarray}
\end{subequations}
where 
 $ D ( q , i \omega ) = 1 
 - 4 f_0^2
  \chi_{\uparrow \uparrow}^{0}
  \chi_{\downarrow \downarrow}^{0}$,
and
 \begin{equation}
 \chi^{0}_{\sigma \sigma^{\prime} } ( q , i \omega ) =
 -  \frac{1}{L} \sum_k \frac{  f (\xi_{ k+q/2,  \sigma^{\prime} } ) -  f ( \xi_{k -q /2, \sigma} ) }{
 \xi_{ k+q/2,  \sigma^{\prime} }  -  \xi_{k -q/2 , \sigma} - i \omega  } 
 \label{eq:chi0tr}
 \; .
 \end{equation}
Here $f ( E )$ is the Fermi function.
For small  $q$ and $\omega$ we may approximate
 \begin{eqnarray}
  \chi^0_{ \sigma \sigma } ( q , i \omega ) & \approx & \frac{ v_{ \sigma}}{\pi} 
 \frac{ q^2}{ ( v_{ \sigma} q )^2 + \omega^2 }
 \; .
 \label{eq:chi0}
 \end{eqnarray}
For $\omega > 0$ 
the dynamic structure factors
$S^{\rm RPA}_i ( q , \omega ) =  \pi^{-1} {\rm Im} \chi_{ii}^{\rm RPA} ( q , \omega + i 0 )$
can then be written as
\begin{equation}
S^{\rm RPA}_i ( q , \omega ) = Z_i | q | \delta ( \omega - v_i | q | ) \;,
\end{equation}
with $ Z_n =  [ \pi \sqrt{  1+ F_0 }]^{-1}$, $v_n = v_F \sqrt{ 1+ F_0 }$, and
$Z_m = [ \pi \sqrt{ \delta_0 }]^{-1}$, $v_m = v_F \sqrt{\delta_0}$.
Note that 
 $S^{\rm RPA}_{i} ( q , \omega )$ satisfy
the sum rules \cite{Forster75}
$ 2 \lim_{ q \rightarrow 0 } \int_0^{\infty} \frac{d \omega}{\omega} \, S^{\rm RPA}_{i} ( q , \omega ) 
  =    \chi_i $,
with the compressibility  $\chi_n = [ \pi v_F ( 1 + F_0  )]^{-1} $ and
the spin susceptibility $\chi_m =  2 / ( \pi v_F \delta_0 )$.
The latter is  related to 
the Hartree-Fock energy (\ref{eq:hartreegain}) via
 $\chi_m^{-1} = L^{-1} 
 \left.  \partial^2 \Omega_0 ( m ) /  \partial { m^2 } 
 \right|_{m_0 }$.
We conclude that
longitudinal spin fluctuations in $1d$ can propagate ballistically, with
velocity $v_m \ll v_F$.
In contrast,  in $3d$ itinerant ferromagnets 
the longitudinal spin mode
can decay into particle-hole pairs and is therefore strongly Landau-damped \cite{Moriya85}.

\begin{figure}[tb]
\begin{center}
\epsfig{file=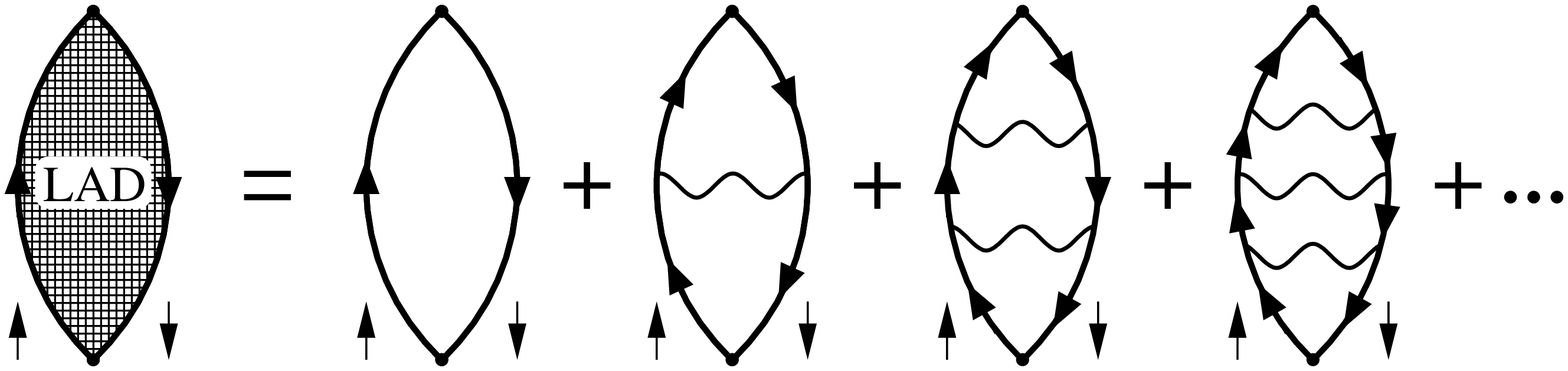,width=80mm}
\end{center}
\vspace{-3mm}
\caption{%
 Ladder approximation for $\chi_{\uparrow \downarrow}$, see Eq.\ (\ref{eq:chibot}).
The solid arrows are the Hartree-Fock single particle Green functions
with given spin projections and the wavy lines
represent the bare interaction.
}
\label{fig:ladder}
\end{figure}

Next, let us calculate the transverse spin-spin correlation function
$\chi_{\uparrow \downarrow}  ( q , i \omega )$ 
within the ladder approximation shown 
in  Fig.\ \ref{fig:ladder}, which yields
\begin{equation}
 \chi_{\uparrow \downarrow}^{\rm LAD}  ( q , i \omega ) =
 [  \chi^{0}_{\uparrow \downarrow } 
( q , i \omega )^{-1} -  2 f_0  ]^{-1}
 \; .
 \label{eq:chibot}
 \end{equation}
For $ | q | \ll q_m $ and $ | \omega|  \ll \Delta$ we may expand
\begin{equation}
\chi^{0}_{\uparrow \downarrow } ( q , i \omega ) \approx 
  \frac{m_0}{ 2 \Delta} \left[ 1 + \frac{ i \omega}{  2 \Delta  } - B q^2\right ] \;,
\label{eq:chibot0}
\end{equation}
with the nonuniversal constant \cite{footnoteB}
$B =  \frac{1}{12} [ \lambda_2 -   \lambda_1^2 ]$. Note that $B \ge A > 0$.
Using
$\Delta = f_0 m_0$ we obtain
 \begin{equation}
 \chi_{\uparrow \downarrow}^{\rm LAD}  ( q , i \omega ) 
=   - \frac{  m_0}{  i \omega -  b q^2  }
 \; ,
 \end{equation}
where $ b = 2 \Delta B$ is the spin wave stiffness.
%
This implies a $\delta$-function peak in the dynamic structure factor,
$S_{\uparrow \downarrow }  ( q , \omega ) = m_0 \delta ( \omega - b q^2 )$, which exhausts
the sum rule $ \int_0^{\infty} \frac{ d \omega}{\omega} \,
S_{\uparrow \downarrow }  ( q , \omega ) = m_0 / b q^2$.
The existence of well-defined transverse spin waves in the
symmetry broken phase follows from general 
hydrodynamic arguments \cite{Forster75}.
However, in $1d$ it may well be that 
interactions lead to anomalous damping of spin waves and a breakdown
of hydrodynamics.
This problem deserves further attention.

Because the ferromagnetic instability is 
triggered by interactions with zero momentum transfer, we expect
that at low energies the relevant interaction is dominated by forward scattering.
Moreover, for $m=0$  it is known that  repulsive
backscattering interactions
are marginally irrelevant \cite{Solyom79}. 
We assume that this remains true in the
ferromagnetic state and expect that
this assumption can be verified using RG methods. 
Note also that  weak ferromagnetism in $3d$ can be understood within the framework
of Fermi liquid theory \cite{Dzyaloshinskii76},
so that it is natural to expect that  Luttinger liquid theory is the corresponding low-energy theory
in  $1d$, at least  if the characteristic magnetic wavevector $q_m $ is
small compared with $k_F$. 

The leading long-distance behavior of correlation functions
can then be obtained from a generalized Tomonaga-Luttinger model, where
the energy dispersion is linearized around the Fermi points $ \pm k_{\sigma}$.
Introducing a bandwidth cutoff $\Lambda$ such that $ q_m \ll \Lambda  \ll k_F$
and defining field operators $ \hat{\psi}^{\alpha}_{ \sigma  } ( q ) = \sqrt{L}
 \, \hat{c}_{ \alpha k_\sigma + q , \sigma }$,
where $ \alpha = \pm 1$ labels the Fermi points,
the kinetic energy is represented by 
$  \sum_{   \alpha \sigma }  \int_{ - \Lambda }^{\Lambda} \frac{d q}{2 \pi} \,
 \alpha v_{\sigma} q  
 \hat{\psi}^{\alpha \dagger}_{ \sigma  } ( q )
 \hat{\psi}^{\alpha}_{ \sigma  } ( q )$.
The interaction is formally identical with Eq.\ (\ref{eq:Hamiltonian}),
but with an implicit momentum transfer cutoff $1/r_0 \ll k_F$ 
and the density operators 
now given by $ \hat{\rho}_n ( q ) = \sum_{\alpha \sigma} \int_{ - \Lambda}^{\Lambda} 
\frac{ d q^{\prime} }{2 \pi } \, \hat{\psi}^{\alpha \dagger}_{ \sigma  } ( q^{\prime} )
 \hat{\psi}^{\alpha}_{ \sigma  } ( q^{\prime} + q )$, and similarly for
the spin-density operator $ \hat{\rho}_m ( q )$.
Moreover, the bare couplings $f_{ij}$
in Eq.\ (\ref{eq:Hamiltonian}) should be replaced by  
renormalized low-energy couplings ${g}_{ij}$, which
characterize the Luttinger liquid fixed point \cite{Penc93}.
Note that for $m \neq 0$ the renormalized interaction
is not spin-rotationally invariant, so that in general
${g}_{nm}  \neq 0$.
However, for $m \ll n $ we expect that the generic behavior
of correlation functions (with the possible exception of 
$\chi_{\uparrow \downarrow } ( q , \omega )$ in the spin wave regime
$ | q | \ll q_m$) can be correctly obtained for  the special case
${g}_{nm}  = 0$ and $ {g}_{nn} = - {g}_{mm} \equiv  {g} > 0 $.

Given the effective low-energy theory,
the closed loop theorem \cite{Dzyaloshinskii73} 
guarantees that all corrections to the RPA for the density-density  and
longitudinal spin-spin
correlation functions cancel for small $q $ and $ \omega $.
Hence
Eqs.\ (\ref{eq:chinn}) and (\ref{eq:chimm}) are asymptotically exact
if we replace the bare quantities
$f_0$, $I_0$ and $\delta_0$  by the corresponding renormalized quantities
 $g$, $I$ and $\delta$. 
In particular, the existence of a propagating longitudinal spin mode with velocity
$v_m = v_F \sqrt{\delta } \ll v_F$ is a robust result, and
not an artifact of the RPA.

Due to the linearized energy dispersion and the irrelevance of
scattering processes with large momentum transfers,
the single-particle Green function $G_{\sigma} ( x , \tau )$ 
can be calculated exactly using bosonization in real space and imaginary time.
For $ {\rm{max} } \{ | x | , v_i |   \tau  | \}  \gg r_0$ we obtain
\begin{eqnarray}
 G_{\sigma}  ( x , \tau ) & = & \frac{1}{ 2 \pi i }
 \left[ \frac{r^2_0}{ x^2 +  v_n^2  \tau^2 } \right]^{{\eta}_n/2 }
 \left[ \frac{r^2_0}{ x^2 +  v_m^2  \tau^2 } \right]^{{\eta}_m/2 }
 \nonumber
 \\
 &  \times & \sum_{\alpha}     
\frac{  e^{ i \alpha k_{\sigma} x }   }{  [ \alpha x +  i v_{n} \tau   ]^{1/2}  
 [ \alpha x + i  v_{m} \tau   ]^{ {1}/{2}} }
 \label{eq:Gres}
\; .
 \end{eqnarray}
Here 
$\eta_i = \frac{1}{4} ( K_i + K_i^{-1} - 2 )$, with
$K_n = [ I + 1 ]^{-1/2}$ and $K_m = [2 ( I-1 )]^{-1/2}$. 
Note that the anomalous dimension
$\eta_m$ of the spin channel {\it{diverges}} for $I \rightarrow 1$. This singularity is also found directly
from the universal Luttinger liquid relation \cite{Schulz90}
$\chi_i =  2 {K_i}/ \pi {v_i}$
together with the above RPA results for $\chi_m$ and $v_m$.
We note that an analogous scenario has recently been found for the
charge channel of the $1d$ $t$-$J$ model in the vicinity of the phase
separation instability \cite{Nakamura97}.

\begin{figure}[tb]
\begin{center}
\epsfig{file=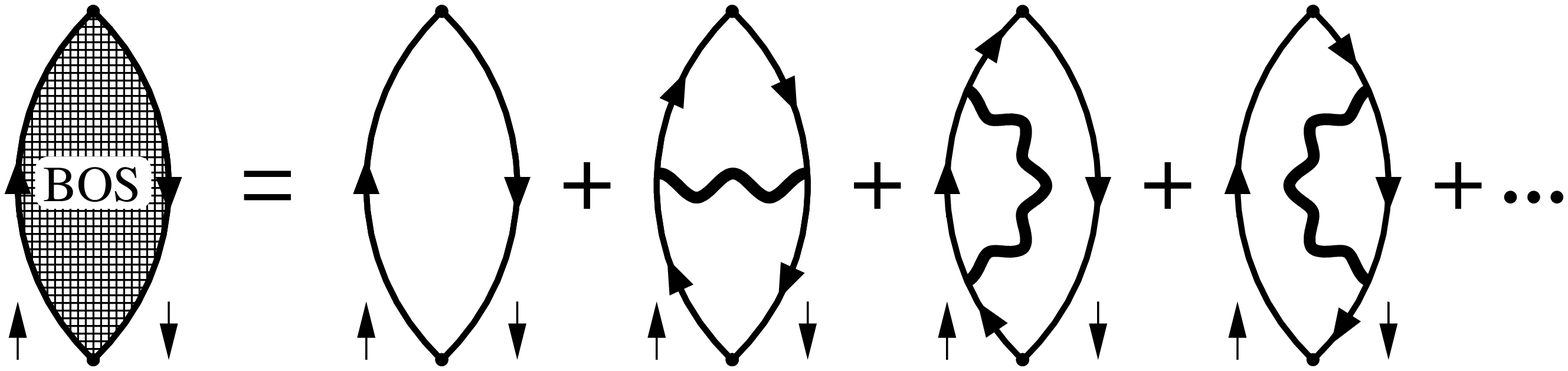,width=80mm}
\end{center}
\vspace{-3mm}
\caption{%
Diagrams contributing to $\chi_{\uparrow \downarrow} $
in bosonization, see
Eq.\ (\ref{eq:chibotres}). The thick wavy line represents the RPA interaction. To second order in the RPA interaction there are already 13 diagrams contributing to $\chi_{\uparrow \downarrow}$.
}
\label{fig:bos}
\end{figure}

\begin{figure}[tb]
\begin{center}
\epsfig{file=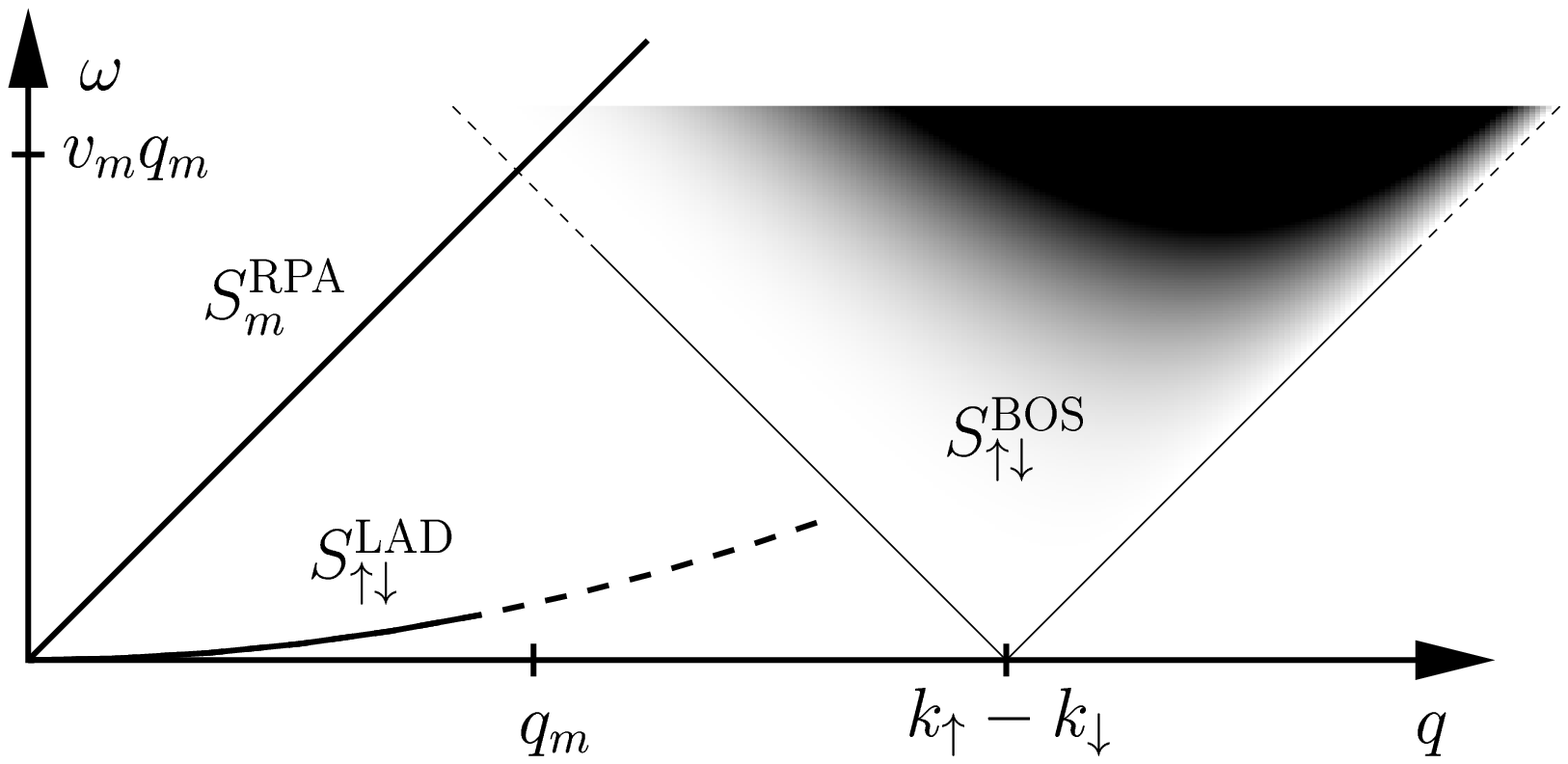,width=80mm}
\end{center}
\vspace{-3mm}
\caption{%
Dispersion of the  longitudinal and the transverse
spin excitations. 
The dashed line indicates that 
only for $ | q | \lesssim q_m$ we expect transverse spin waves to be well-defined.
The triangle touching the horizontal axis at $k_{\uparrow} - k_{\downarrow}$ 
 is the regime where the bosonization result   (\ref{eq:Strans})   for  
$S_{\uparrow \downarrow}  ( q ,  \omega  )$ can be trusted and yields
a finite weight.
The intensity of the shading is proportional to the magnitude of
 $S_{\uparrow \downarrow } ( q , \omega ) $.
}
\label{fig:Stoner}
\end{figure}

Finally, let us consider the transverse spin-spin correlation function
 $ \chi_{\uparrow \downarrow}  ( x , \tau   )$,
which,  due to the linearized energy dispersion, can also be calculated 
for large 
 $x$ and $\tau$ by bosonization. 
Following  Ref.~\onlinecite{Bartosch99} we obtain 
 for $ {\rm{max}} \{ | x |, v_m | \tau |  \}   \gg r_0$ 
\begin{equation}
  \chi_{\uparrow \downarrow}  ( x , \tau  ) =
  \frac{-1}{ (2 \pi )^2}  
 \left[ \frac{ r_0^2 }{  x^2 + v_m^2 \tau^2  } \right]^{ 2 \eta_m } 
\sum_{ \alpha  } 
 \frac{e^{ i  \alpha (k_{\uparrow} - k_{\downarrow} ) x } }{
 [ \alpha x + i v_{  m } \tau ]^2 }
 \; .
 \label{eq:chibotres}
 \end{equation}
The leading diagrams taken into account  in Eq.\ (\ref{eq:chibotres}) are
shown in Fig.\ \ref{fig:bos};
they contain  the ladder diagrams of
Fig.\ \ref{fig:ladder} as a subset, but
include in addition self-energy corrections, screening bubbles,  and complicated vertex corrections.
It is important to realize that Eq.\ (\ref{eq:chibotres}) 
can only be used
to obtain the Fourier transform
$ \chi_{\uparrow \downarrow}  ( q , i \omega  )  =  \int dx d \tau \,
 e^{ - i ( q x - \omega \tau )}  \chi_{\uparrow \downarrow}  ( x , \tau )$
for wavevectors close to 
$ \pm ( k_{\uparrow } - k_{\downarrow } ) $, i.e. for
$ | q \mp ( k_{\uparrow } - k_{\downarrow } ) | \lesssim q_m$.
In the spin wave regime $ |q| \ll  q_m$
the transverse spin-spin  correlation
function cannot be calculated using abelian bosonization
with linearized energy dispersion, because
(i) the ladder approximation suggests that
the spin wave dispersion depends on the nonlinear terms of the energy dispersion, and
(ii) the existence of spin waves follows from the  {\it{spontaneous}} breaking
of spin-rotational invariance, so that their dispersion cannot be obtained 
using a method  which
{\it{explicitly}} violates this symmetry.
On the other hand, for
$ | q \mp ( k_{\uparrow} - k_{\downarrow} ) | \lesssim q_m$
the  Fourier transform
of Eq.\ (\ref{eq:chibotres})  yields an accurate approximation for the
transverse dynamic structure factor
$ S_{\uparrow \downarrow } ( q , \omega )$.
For $ \omega > 0$ we obtain
 \begin{align}
 S_{\uparrow \downarrow } ( q , \omega )  = & \,
 C_m  
\Theta ( \omega-v_m ||q| - k_{\uparrow} + k_{\downarrow} | ) \nonumber \\
 & \hspace{0mm} \times [ \omega-v_m (|q| - k_{\uparrow}+k_{\downarrow}) ]^{2\eta_m-1}  \nonumber \\
 & \hspace{0mm} \times 
 [\omega+v_m ( |q| - k_{\uparrow} + k_{\downarrow})]^{2\eta_m+1} \;,
 \label{eq:Strans}
 \end{align}
with $C_m=[4\pi v_m 
\Gamma(2\eta_m)\Gamma(2+2\eta_m)]^{-1} ({r_0}/{v_m})^{4\eta_m}$.
The region where  $S_{\uparrow \downarrow } ( q , \omega ) $ is finite
represents the $1d$ Stoner continuum. The complete picture of low-energy spin excitations is depicted in Fig.\ \ref{fig:Stoner} and is qualitatively quite similar to its $3d$ counterpart \cite{Moriya85}.
However, in $1d$ there is no Landau damping and the structure factor shows anomalous scaling associated with broken spin-rotational symmetry of a Luttinger liquid phase.

For an outlook from a renormalization-group perspective, we note that
while the ferromagnetic ground state is stabilized by nonlinear terms
in the energy dispersion close to the Fermi points, the flow of the
corresponding irrelevant couplings is not accessible within the usual
field-theoretical RG \cite{Solyom79}.  However, using modern
formulations of the RG \cite{Kopietz01} based on Wilson's idea of
eliminating degrees of freedom and rescaling, it should be possible to
examine the subtle role played by irrelevant couplings in
stabilizing a ferromagnetic ground state in $1d$.

In conclusion, we presented the effective low-energy theory of weakly
ferromagnetic Luttinger liquids. Many of their properties only depend
on the effective Stoner parameter $I$, i.e., on the distance
$\delta=2(I-1)/I\ll1$ from the ferromagnetic instability.
%
Neutron scattering experiments should be able to test our predictions for spin-spin correlation functions.
Furthermore the propagating longitudinal mode with small velocity $v_m  \propto \delta^{1/2}$ 
and large residue $Z_m \propto \delta^{-1/2}$ 
dominates some thermodynamic quantities, for example through the divergence
of the uniform spin susceptibility,
$ \chi_m  \propto Z_m / v_m \propto \delta^{-1}$.
The discussed features of the weakly ferromagnetic regime should be accessible
in specially designed organic polymers \cite{Arita02}, for which  the effective Stoner parameter $I$ can be controlled by adjusting the
density via external gate voltages. 
Our predictions are also relevant to semiconductor quantum wires which are believed to show spontaneous ferromagnetism \cite{Thomas96,Reilly02,PhysicsToday02}.

This work was supported by the DFG via Forschergruppe FOR 412, Project No. KO 1442/5-1.


\begin{thebibliography}{99}
\bibitem{footnoteHubbard}
In the Hubbard model
$f_{n}  = - f_{m}  = Ua/2$, where $U$ is the on-site interaction and
 $a$ is the lattice spacing.
\end{thebibliography}
\end{document}